\documentclass[prper,aps,twocolumn,groupedaddress,floats,showpacs,final,superscriptaddress]{revtex4-1}
\usepackage{leading}
\usepackage[latin3]{inputenc}
\usepackage[makeroom]{cancel}
\usepackage{graphicx}
\usepackage{amsmath}
\usepackage{amsfonts}
\usepackage{amssymb}

\usepackage{color}
\usepackage[left=2cm,right=2cm,top=2cm,bottom=2cm]{geometry}
\usepackage{graphicx} 
\usepackage{dcolumn}
\usepackage{bm}
\usepackage{simplewick}
\usepackage{array}
\usepackage{appendix}

\RequirePackage[
   hyperindex,colorlinks,bookmarksnumbered,
   plainpages=true,pdfstartview=FitH]{hyperref}
\hypersetup{linkcolor=blue,urlcolor=blue,citecolor=blue}
\usepackage{hyperref}
\usepackage{mathrsfs}
\definecolor{purple}{rgb}{0.5,0,0.6}

\usepackage{ulem}
\renewcommand{\emph}[1]{\textit{#1}}
\definecolor{darkblue}{rgb}{0,0,0.5}
\definecolor{darkgreen}{rgb}{0,0.5,0}
\definecolor{darkred}{rgb}{.7,0,0}
\definecolor{purple}{rgb}{0.5,0,0.6}
\definecolor{orange}{rgb}{1,0.5,0}
\definecolor{grey}{rgb}{.6,.6,.6}
\definecolor{lightpink}{rgb}{1,0.7,0.75}
\definecolor{pink}{rgb}{1,0.4,0.58}
\definecolor{deeppink}{rgb}{1,0.08,0.58}

\newcommand{\DK}[1]{{\color{black}{#1}}} 
\newcommand{\Dk}[1]{{\color{black}{#1}}} 

\renewcommand{\emph}[1]{\textit{#1}}

\newcommand{\mt}{\mathcal{T}}
\newcommand{\ec}{E_{\rm C}}

\begin{document}

\title{Coulomb blockade oscillations of heat conductance in the charge Kondo regime}

\author{D. B. Karki}
\affiliation{Division of Quantum State of Matter, Beijing Academy of Quantum Information Sciences, Beijing 100193, China}

\begin{abstract}
We develop a method of theoretically investigating the charge, energy and heat transport in the presence of the charge Kondo correlations.  The Coulomb blockade oscillations of heat conductance in the single electron transistor exhibiting the charge Kondo effects are investigated. We explore the Wiedemann Franz ratio in both charge-single channel and charge-two channel Kondo regime. The close connections of our findings with the recent experiments on multi-channel charge Kondo effects are discussed.
\end{abstract}

\date{\today}
\maketitle

\section{Introduction} 
\vspace*{-3mm}
The rapid progress of the quantum technologies has led to the new experiments in addition to the development of new theoretical approaches to understand the quantum transport phenomena at the nanoscale~\cite{casti}. A prototypical nanoscale device of common interests is a single electron transistor (SET) which contains a metallic grain, the quantum dot (QD), tunnel coupled to two electronic reservoirs~\cite{ld0, casti}. The transport characterization of such a SET is mainly governed by the Coulomb blockade (CB) phenomena~\cite{ben}. Consequently, the electronic conductance and the thermopower of a SET show the CB oscillations as a function of the gate voltage as reported in the seminal work~\cite{Staring_1993}. The theory of CB oscillation based on the sequential tunneling approach originally developed in Ref.~\cite{ben}, however, does not consistently explains the low temperature transport characteristics of a SET since the corresponding transport is dominated by the inelastic cotunneling mechanisms~\cite{cbo1, lm1}.

The complete theory of low temperature thermoelectrics of a SET came after the seminal works~\cite{matv1, matv2, matv3, matv4, matv5}. These perseverances have investigated the CB oscillations of thermoelectric coefficients of a SET with the QD strongly coupled to one of the leads by a quantum point contact as in the experiment reported in Ref.~\cite{lm1}. Interestingly, the SET setup with the QD strongly coupled to one of the leads has been shown to be associated with the single channel Kondo (1CK) and two channel Kondo (2CK) effects~\cite{matv1}. The Kondo effect involving the Coulomb blockade in QD arises from the two possible charge states in the QD which are adjusted to the same energy by tuning the gate voltage to the critical point~\cite{matv2}. These two charge states then behave fundamentally as same as the effective spin eigenstates and hence the electron spin plays the role of two Kondo channels~\cite{affjs}. Therefore, the spin polarization by magnetic field results in the charge-1CK effects and the electrons with an effective spin-1/2 provide the realization of charge-2CK effects in the original formulation~\cite{matv1}. 

The Kondo paradigm provides a valuable tool of understanding the physics of strongly correlated systems both in Fermi-liquid (FL) and non-FL (NFL) regime~\cite{kondo, Anderson, haman1,  Anderson_Yuval_Hamann,Nozieres,Wilson,Nozieres_Blandin_JPhys_1980,wigmann_JETP(38)_1983,jvm1, AFFLECK_NPB_1990,Affleck_Lud_PRB(48)_1993}. \DK{So far much progresses have been made for the understanding of spin Kondo effects that stem from a localized spin at a discrete energy level in QD nanostructures~\cite{revival}. Although both the charge Kondo effects and the spin Kondo effects can be observed in the QD, their behaviors and the corresponding fundamental mechanisms are strikingly different~\cite{affjs}. Consequently, one expects the different scaling behaviors of transport properties in the spin and charge versions of the Kondo effects. }

The charge-1CK falls into the FL universality class the description of charge-2CK is beyond the scope of FL theory~\cite{Nozieres,Nozieres_Blandin_JPhys_1980}. Therefore, the theoretical advancements made in Refs.~\cite{matv1, matv2, matv3, matv4, matv5} also provide a viable root of experimentally accessing the NFL regime as reported in recent experiments on the charge-2CK effect~\cite{Pierre_NAT(526)_2015}. In addition, since the first proposal~\cite{matv1}, various transport properties in charge-1CK and charge-2CK regime have been intensively investigated by different theoretical methods~\cite{matv2, matv3, matv4, matv5,tkt1, tkt2, ss1, tkt3}. In the recent years, the interests in charge-2CK have been also expanded to the study of energy and heat transports~\cite{mk, mkk} which can be measured with the existing experimental setups~\cite{Pierre_NAT(526)_2015, iff}.

For a QD based SET in the presence of the voltage bias and temperature gradient, various measure of thermoelectric response have been the subject of recent experimental and theoretical perseverances~\cite{casti}. It is almost two decades since the development of a full-fledged theory of thermopower in charge-2CK regime of a QD~\cite{matv3}. The original theory has been then extended to the more complex geometries exhibiting charge-2CK effects~\cite{tkt1, tkt2}. However, very less attentions have been paid for the investigation of heat response in charge-2CK regime although it is within the experimental access~\cite{Pierre_NAT(526)_2015, mkk}. In this work, we aim to fill this gap by developing a general theoretical framework dealing with the charge, energy and heat transport in the charge Kondo regime based on the original proposal~\cite{matv3}. \DK{In particular, we focus mainly on the aspects of heat transport in the presence of charge Kondo correlation while the corresponding charge and energy transport have been throughly investigated in Refs.~\cite{matv3, matv4}.}

The paper is organized as follows. In Sec.~II, we discuss briefly on the common measures of 
thermoelectric transport including the electronic conductance $G$, thermopower $S$ (Seebeck coefficient) and electronic thermal conductance $\mathcal{K}$. In addition, we present the connection between $G$ and $S$ as predicted by the semi-classical Cutler-Mott relation. We also discuss the impact of Kondo correlation on the Wiedemann-Franz ratio connecting $G$, $S$ and $\mathcal{K}$. We present the model description and an outline of the calculation of transmission coefficient, the fundamental quantity characterizing the transport of charge, energy and heat, of a QD based SET in Sec.~III. The Sec.~IV is devoted to the investigation of the charge, energy and heat transport in case of spinless electrons exhibiting the charge-1CK effects. The transport in charge-2CK regime is explored in Sec.~V. The last section~VI contains the conclusion of our work together with the possible future
research plans based on the present work.

\section{Thermoelectric transport coefficients}
\vspace*{-3mm}
The general setup for the thermoelectric transport contains the quantum impurity (QD) tunnel coupled to two external electron reservoirs. The left (L) and right (R) reservoirs are in equilibrium, separately, at temperatures $T_{\gamma}$ ($\gamma{=}\rm {L, R}$) and chemical potentials $\mu_{\gamma}$ respectively. Once the temperature gradient $\Delta T\equiv T_{\rm L}{-} T_{\rm R}$ and the voltage bias $e\Delta V\equiv\mu_{\rm L}{-} \mu_{\rm R}$ are established across the setup, the heat current ($I_{\rm h}$) and charge current ($I_{\rm c}$) start to flow. For simplicity of the presentation, henceforth we use the system of atomic unit $e=k_{\rm B}=\hbar=1$. The charge and the heat currents in the linear response theory are then connected by the Onsager relations~\cite{sagar1, sagar2}, 
\begin{equation}\label{ym7}
 \left(%
\begin{array}{c}
  I_{\rm c} \\
  I_{\rm h} \\
\end{array}%
\right)= \left(%
\begin{array}{cc}
  {\rm L}_{11} & {\rm L}_{12} \\
  {\rm L}_{21} & {\rm L}_{22} \\
\end{array}%
\right)\left(%
\begin{array}{c}
  \Delta V \\
  \Delta T \\
\end{array}%
\right).
\end{equation}
The Onsager transport coefficients ${\rm L_{\rm ij}}$ in Eq.~\eqref{ym7} provide all the thermoelectric measurements of interests in linear response regime~\cite{costi1}. To further deal with the coefficients ${\rm L_{\rm ij}}$, we setup the transport integrals in terms of the transmission coefficient $\mt(\varepsilon, T)$~\cite{Zlatic, dee1, last1, last2}
\begin{equation}\label{ym8}
\mathscr{L}_{\rm n}\equiv \frac{1}{4T}\int^{\infty}_{-\infty}d\varepsilon\;\frac{\varepsilon^n}{\cosh ^2\left(\frac{\varepsilon}{2 T}\right)}\;\mt(\varepsilon, T),\;\;{\rm n}=0, 1, 2,
\end{equation}
where $T$ is the reference temperature. 

The transport integrals $\mathscr{L}_{\rm n}$ are then directly connected to the Onsager transport coefficients ${\rm L_{\rm ij}}$, namely ${\rm L_{11}}=\mathscr{L}_0$ and ${\rm L_{12}}=\mathscr{L}_1/T$~\cite{kim}. In addition ${\rm L_{12}}$ and ${\rm L_{21}}$ are related by the Onsager reciprocity relation and the coefficient ${\rm L_{22}}$ relates the electronic thermal conductance with ${\rm L_{11}}$ and ${\rm L_{12}}$~\cite{casti}. While the electronic conductance directly follows from $\mathscr{L}_0$, the thermopower or the Seebeck coefficient is usually obtained as $S=\mathscr{L}_1/\mathscr{L}_0 T$. In case of the noninteracting electrons in a metal, the Cutler-Mott relation~\cite{cutler} connects the thermopower $S_{\rm CM}$ with the logarithmic derivative of the energy dependent electronic conductance $G(E)$ with respect to the energy $E$
\begin{equation}\label{mmh}
S_{\rm CM}=\frac{\pi^2}{3}T\;\frac{d\ln G(E)}{d E}.
\end{equation}
The deviation of $S$ from $S_{\rm CM}$ amounts to the strong electron correlation in the system~\cite{kiselev}. In addition to the Seebeck effect, the Peltier effects are also of the common interests in the generic thermoelectric experiments~\cite{casti}. The Peltier effect describes the generation of a
heat current $I_{\rm h}$ due to the charge current $I_{\rm c}$ driven in
a circuit under isothermal condition $T_{\rm L}{=}T_{\rm R}$ by an applied voltage bias $\Delta V$. The Peltier coefficient $\Pi^{\rm \gamma}$ associated with the $\gamma$ reservoir is defined as: $\Pi^{\rm \gamma}=\left.I^{\rm \gamma}_{\rm h}/I_{\rm c}\right|_{T_{\rm L}{=}T_{\rm R}}$~\cite{landm1}. This coefficient provides the valuable informations on the characterization of how
good a material is for thermoelectric solid-state refrigeration
or power generation~\cite{casti}. In addition, linear response (LR) Peltier coefficient $\Pi_0$ is related to the corresponding Seebeck coefficient $S=S^{\rm LR}$ via the Kelvin relation $\Pi_0=TS$~\cite{kim}.

The investigations beyond the thermopower are usually done by the electronic thermal conductance $\mathcal{K}$. The Wiedemann-Franz (WF) law connects the electronic thermal conductance $\mathcal{K}$ to the electrical conductance $G$ in low temperature regime of a macroscopic sample by an universal constant, the Lorenz number $L_0$, defined as $L_0\equiv \mathcal{K}/GT=\pi^2/3$~\cite{Zlatic, casti}. However, the transport through the nano devices generally expected to violate the WF law even at the FL regime~\cite{casti}. Interestingly, the WF law has been recently reported to be satisfied even at the NFL regime of Kondo effects~\cite{dbk, mkk}. The violation or the validation of WF law is usually accounted for by
studying the Lorenz ratio $R$ which is expressed in terms of the transport integrals~\cite{lastdd}
\begin{equation}\label{ym9}
R(T)\equiv\frac{L(T)}{L_0}=\frac{3}{(\pi T)^2}\left[\frac{\mathscr{L}_{\rm 2}}{\mathscr{L}_{\rm 0}}-\left(\frac{\mathscr{L}_{\rm 1}}{\mathscr{L}_{\rm 0}}\right)^2\right],
\end{equation}
any deviation of $R$ from unity, therefore, amounts the violation of WF law.

From the preceding discussions, it is apparent that the fundamental quantities of characterizing the LR transport properties of a SET considered in this work are the transport integrals $\mathscr{L}_{\rm n}$. Furthermore, $\mathscr{L}_{\rm n}$ are the function of the transmission coefficient $\mt(\varepsilon, T)$ which depends on the detail of the model. In the following section, we discuss in detail about the calculation of the transmission coefficient applicable to both the charge-1CK and charge-2CK regime of a QD based SET.

\section{Model Hamiltonian}
\vspace*{-3mm}
We consider a QD (metallic island) coupled to the two electronic reservoirs, the left (L) lead and the right (R) lead. While the coupling between the QD and left lead is provided by a tunneling junction, that with the right lead is achieved by a single channel quantum point contact (QPC) with reflection amplitude $|r|$ (see Ref.~\cite{matv1} for details). \DK{In addition, the conductance of the tunneling junction connecting the QD to left lead $G_{\rm L}$ is assumed to be much smaller compared to that of the other junction $G_{\rm R}$. This results in the thermal equilibrium between the QD and the right lead. For this case, namely the QD coupled weakly to the left contact and strongly to the right contact, the explicit form of low energy Hamiltonian  $\mathscr{H}=\mathscr{H}_0+\mathscr{H}_{\rm L}+\mathscr{H}_{\rm R}+\mathscr{H}_{\rm C}$ well known in the language of bosonization~\cite{matv3} (for the clarity of presentation, in the following we write the Hamiltonian from the Ref.~\cite{matv3}). Here the noninteracting part of the Hamiltonian reads~\cite{matv3, matv4}
\begin{align}
\mathscr{H}_{0}&=\sum_{k\sigma}\varepsilon_{k\sigma}c_{k\sigma}^{\dagger}c_{k\sigma}+
\sum_{\sigma}\varepsilon_{\sigma}d_{\sigma}^{\dagger}d_{\sigma}\nonumber\\
&+\frac{v_{\rm F}}{2\pi}\sum_\sigma\int^{\infty}_{\infty}\!\Big(\pi^2 \Pi_\sigma^2(x)+[\partial_x \phi_\sigma(x)]^2\Big)dx,
\end{align}
where the operator $c_{k\sigma}$ annihilates an electron in the momentum state $k$ with spin $\sigma=\uparrow, \downarrow$ in the left lead, $d_\sigma$ annihilates an electron of spin $\sigma$ in the QD and $v_{\rm F}$ stands for the Fermi velocity. The bosonized displacement operator $\phi_\sigma$ and corresponding conjugated momentum operator $\Pi_\sigma$ satisfy the usual commutation relation connecting the $\delta$-function: $[\phi(x), \Pi(y)]=i\delta(x-y)$. 

The Hamiltonian $\mathscr{H}_{\rm L}$ describing the tunneling from the left lead to the QD is given by\Dk{
\begin{equation}
\mathscr{H}_{\rm L}=\sum_{k\sigma}\left(t_{k}c^{\dagger}_{k\sigma}d_{\sigma}+{\rm h.c.}\right),
\end{equation}
with $t_{k}$} being the tunneling amplitude. The backscattering in the QPC is accounted for by the Hamiltonian 
\begin{equation}
\mathscr{H}_{\rm R}=-\frac{D}{\pi}|r|\sum_\sigma\cos[2\phi_\sigma(0)],
\end{equation}
where $D$ is the bandwidth. In addition, the Coulomb interaction in the QD is described by the Hamiltonian
\begin{equation}\label{bder2}
\mathscr{H}_{\rm C}=\ec \Big(\hat{n}+\frac{1}{\pi}\sum_\sigma\phi(0)-N\Big)^2.
\end{equation}
\Dk{Here $\hat{n}$ is the integer valued operator that commutes with the electron annihilation operator of the left contact $\psi_{\rm L}$, $N$ is a dimensionless parameter proportional to the gate voltage and $\ec$ is the charging energy of island (see Ref.~\cite{matv4} for details).}}
\section{Transmission coefficient at charge Kondo regime}
\DK{In the following we describe the tunneling through the left contact within the second order of perturbation in the corresponding tunneling matrix element \Dk{$t_{k}$}. Under this assumption the electron transport processes are explained} in terms of the tunneling density of states (DOS) of the left lead $\nu_{\rm L}(\varepsilon)$ and that of the QD $\nu_{\rm D}(\varepsilon)$~\cite{matv3}. In addition, due to the weak energy dependence, henceforth we consider the case of  $\nu_{\rm L}(\varepsilon)=\nu_{\rm L}$. We \Dk{then} define the finite temperature $T$ transmission coefficient $\mt(\varepsilon, T)$ characterizing the charge, energy and heat transport through the left contact by the following relation~\cite{matv1, matv2} 
\begin{equation}\label{ym3}
\mt(\varepsilon, T)=-\frac{G_{\rm L}}{\nu_0}\;\nu_{\rm D}(\varepsilon, T).
\end{equation}
In Eq.~\eqref{ym3}, $G_{\rm L}\equiv 2\pi \nu_{\rm L}\nu_0\langle|t_{kk'}|^2\rangle$ is the conductance of the left barrier for noninteracting electrons in the QD and $\nu_0$ stands for the DOS in the QD which is no longer renormalized by the electron interactions. The tunneling DOS of the QD is then given in terms of electron Green's function (GF)~\cite{matv1, matv2}, 
\begin{equation}\label{ym4}
\nu_{\rm D}(\varepsilon, T)={-}\frac{1}{\pi}\!\cosh\left(\frac{\varepsilon}{2 T}\right)\!\int^{\infty}_{-\infty}\!\!\mathcal{G}\left(\frac{1}{2 T}{+}it\right)e^{i\varepsilon t}\;dt,
\end{equation}
where $\mathcal{G}(\tau){=}{-}\langle T_{\tau} \psi_{\rm L}(\tau)\psi^{\dagger}_{\rm L}(0)\rangle$ is the Matsubara GF defined in terms of the operator $\psi_{\rm L}$ which annihilates an electron in the QD at the position of the left contact~\cite{matv2}.

\DK{Rescaling the operator $\psi_{\rm L}\equiv \psi_{\rm L}F$ in terms of another operator $F$ which lowers $\hat{n}$ by unity (that means the commutation relation $[F, \hat{n}]=F$ to be satisfied), we define} the GF $\mathcal{G}(\tau)$ given in Eq.~\eqref{ym4} as the product of noninteracting part $\mathcal{G}_0(\tau)$ and the correlator $K(\tau)$ accounting for the electron interactions in the system~\cite{matv1, matv2, matv3, matv4, matv5} 
\begin{equation}\label{ym5}
\mathcal{G}(\tau)=\mathcal{G}_0(\tau)\;K(\tau)=-\frac{\pi\nu_0 T}{\sin(\pi T \tau)}\;K(\tau).
\end{equation}
\DK{Here the time order correlator $K(\tau)$ is defined in terms of $F(\tau)$ such that $K(\tau)\equiv \langle T_\tau F(\tau) F^{\dagger}(\tau)\rangle$ with $T_\tau$ being the corresponding time ordering operator.} 

Plugging in the transmission coefficient given by Eq.~\eqref{ym3} into the transport integrals Eq.~\eqref{ym8} followed by the use of exact integrals calculated by the method of contour integration
\begin{align}
\mathcal{I}_n&\equiv\int^{\infty}_{-\infty} \frac{x^n\;e^{i2 T t x}}{\cosh x}\;dx,\;{\rm with}\;\;\mathcal{I}_0=\frac{\pi}{\cosh(\pi T t)},\nonumber\\
\mathcal{I}_1&{=}i\frac{\pi^2}{2}\frac{\sinh(\pi T t)}{\cosh^2(\pi T t)},\;\;
\mathcal{I}_2{=}\frac{\pi^2}{4}\Big[\frac{2\pi}{\cosh^3(\pi T t)}{-}\mathcal{I}_0\Big],
\end{align}
we obtain the most general form of the transport coefficients in terms of the electron correlator $K(\tau)$
\begin{align}\label{mmj}
&\mathscr{L}_{\rm 0}=-\pi T\frac{G_{\rm L}}{2}\int^{\infty}_{-\infty}\!\!\!\!dt\;K\left(\frac{1}{2 T}{+}it\right)\;\frac{1}{\cosh^2(\pi T t)},\nonumber\\
&\mathscr{L}_{\rm 1}={-}i(\pi T)^2\frac{G_{\rm L}}{2}\int^{\infty}_{-\infty}\!\!\!\!dt\;K\left(\frac{1}{2 T}{+}it\right)\;\frac{\sinh(\pi T t)}{\cosh^3(\pi T t)},\nonumber\\
&\mathscr{L}_{\rm 2}{=}{-}(\pi T)^3\frac{G_{\rm L}}{2}\!\!\!\int^{\infty}_{-\infty}\!\!\!\!\!\!dt K\left(\!\!\frac{1}{2 T}{+}it\!\right)\!\frac{2}{\cosh^4(\pi T t)}{-}(\pi T)^2\mathscr{L}_0.
\end{align}
\DK{Computation of thermoelectric transport coefficients in Eq.~\eqref{mmj} essentially needs the explicit form of electron correlator $K(\tau)$.} In addition, $K(\tau)$ also depends on the numbers of the conduction channels in the contact which connects the left lead and the QD. For our propose of investigating the single channel and two channel charge Kondo effects, it is however sufficient to consider a single-channel contact connecting the left lead and the QD. Furthermore, since the electron spin behaves fundamentally as same as the Kondo channels, the transport mechanism for spinless electrons and that for the spin-1/2 electrons are strikingly different. In the following we discuss case of spinless electrons and that with spin-1/2 separately.

\section{The charge-1CK regime: spinless electrons}
\vspace*{-3mm}
In case of spinless electrons which is generally achieved by applying the magnetic field, the SET setup exhibits the charge-1CK effect. The time order correlator in charge-1CK regime has been shown to possess a compact analytical expression~\cite{matv4}
\begin{align}\label{aamoi1}
K\left(\!\!\frac{1}{2T}{+}i t\!\right)&{=}\!\left(\frac{\pi^2 T}{\gamma\ec}\right)^2\!\!\!\!\frac{1}{\cosh^2(\pi T t)}\!\Big[\!1{-}2\gamma\xi |r|\!\cos(2\pi N)\nonumber\\
&{-}i4\pi^2\xi\gamma|r|\frac{T}{\ec}\sin(2\pi N)\tanh(\pi Tt)\Big],
\end{align}
where $\xi\simeq 1.59$ is a constant and the symbol $\gamma$ is related to the Euler constant $C$ such that $\ln\gamma=C$. In addition, we explicitly assumed the low-temperature regime satisfying the condition $T\ll\ec$. Substitution of the correlator Eq.~\eqref{aamoi1} into Eq.~\eqref{mmj} results in the following expressions of the transport integrals
\begin{align}\label{aamoi2}
\mathscr{L}_{\rm 0}&=\frac{2}{3}G_{\rm L}\left(\frac{\pi^2 T}{\gamma\ec}\right)^2\Big[1-2\gamma\xi |r|\cos(2\pi N)\Big], \nonumber\\
\mathscr{L}_{\rm 1}&=-\frac{8\pi^7\xi G_{\rm L}}{15\gamma}T\left(\frac{T}{\ec}\right)^3 |r|\sin(2 \pi N),\nonumber\\
\mathscr{L}_2&=\frac{2}{5}G_{\rm L}(\pi T)^2\!\!\left(\frac{\pi^2 T}{\gamma\ec}\right)^2\!\!\Big[1{-}2\gamma\xi |r|\cos(2\pi N)\Big].
\end{align}
To obtain the compact form of the transport integrals Eq.~\eqref{aamoi2}, we integrated out the $t$ variable by the method of contour integration. Namely, we used the integrals of the form
\begin{align}
\mathscr{O}_n\equiv\int^{\infty}_{-\infty}\!\!\!\!dt\;\frac{e^{i a t}}{\cosh^n(\pi T t)},
\end{align}
for $n=4$ and $n=6$
\begin{align}
&\mathscr{O}_4=-\frac{a\pi}{6(\pi T)^4}\frac{ \left(a^2+4 \pi ^2 T^2\right) }{\sinh(a/2T)},\nonumber\\
&\mathscr{O}_6={-}\frac{a\pi}{120(\pi T)^6}\frac{\left(a^4{+}20 \pi ^2 a^2 T^2{+}64 \pi ^4 T^4\right)}{\sinh(a/2 T)}.
\end{align}

The transport coefficients given by Eq.~\eqref{aamoi2} result in the expressions of the thermopower and the Lorenz ratio
\begin{align}\label{bjy1}
S &=-\frac{4\pi^3\xi\gamma}{5\ec}|r|\sin(2\pi N)\;T,\\
&\;\;\;R=\frac{9}{5}-\frac{3}{\pi^2}\;S^2.
\end{align}
To obtain the thermopower expression Eq.~\eqref{bjy1}, we used only the leading term in the conductance. \DK{We note that the expression of thermopower presented in Eq.~\eqref{aamoi2} have been already established in Ref.~\cite{matv3}}. The quadratic temperature scaling of the electronic conductance $\mathscr{L}_0$ and the linear suppression of thermopower with temperature as well as the reflection coefficient as seen from the Eqs.~\eqref{aamoi1} and~\eqref{bjy1} signifies the FL behavior as similar to the single channel spin Kondo effects. \Dk{Moreover, at the particle-hole (PH) symmetric point ($S=0$) the Lorenz ratio attains an universal constant $9/5$ even at finite temperature~\footnote{Note that the original proposal in Ref.~\cite{matv3} is limited to the temperature which should be larger than the tunnel coupling strength to the left lead. Therefore, exactly zero temperature limit can not be reached in our work.}}.

In addition from the Lorenz ratio given in Eq.~\eqref{bjy1}, one can analyze the low temperature behavior of the electronic thermal conductance $\mathcal{K}\propto\mathscr{L}_0 RT$. Since $G\propto T^2$ and $S\propto T$, the leading temperature scaling of the conductance is $\mathcal{K}\propto T^3$. This cubic temperature scaling of $\mathcal{K}$ is in contrast to the single channel spin Kondo effects where the leading temperature scaling behavior of $\mathcal{K}$ is linear due to the presence of unitary conductance.
\section{The charge-2CK regime: electrons with spin}
\vspace*{-3mm}
The SET setup with electrons having an effective spin-1/2 exhibits the charge-2CK effect at low temperature regime $T\ll \ec$. The time order correlator accounting for the interactions for the case of spin-1/2 electrons \DK{as computed in Ref.~\cite{matv3}} is given by
\begin{align}\label{ym6}
&K\!\left(\!\frac{1}{2T}{+}i t\!\right)\!{=}\frac{\pi\Gamma T}{2\gamma \ec}\mathcal{I}_0{-}\frac{2 T}{\ec}|r|^2 \sin(2 \pi N) \ln\left(\!\!\frac{\ec}{\Gamma{+}T}\right)\!\mathcal{I}_1,\nonumber\\
&\;\;\;\;\;\;\mathcal{I}_n=\frac{1}{\cosh(\pi T t)}\int^{\infty}_{-\infty} dx\;\frac{x^n}{x^2{+}\Gamma^2}\frac{e^{i tx}}{\cosh(x/2T)},
\end{align}
where we defined the gate voltage dependent parameter $\Gamma\equiv 8\gamma\ec |r|^2\cos^2(\pi N)/\pi^2$. By substituting the Eq.~\eqref{ym6} into Eq.~\eqref{mmj}, we obtained the following expressions of the transport integrals
\begin{align}\label{ym10}
\mathscr{L}_0&=\frac{\Gamma G_{\rm L}}{8\gamma \ec}\;\mathcal{P}_0,\;\;\;\mathscr{L}_{\rm 2}=\frac{G_{\rm L}\Gamma}{48\gamma \ec}T^2\mathcal{J}\nonumber\\
\mathscr{L}_1 &=-\frac{ G_{\rm L}}{6\pi} \frac{T^2}{\ec}|r|^2 \sin(2 \pi N) \ln\left(\frac{\ec}{\Gamma+T}\right)\mathcal{P}_1.
\end{align}
The coefficients appearing in Eq.~\eqref{ym10} are expressed into the form of integrals
\begin{align}\label{nn1}
\mathcal{P}_n&=\!\!\int^{\infty}_{-\infty}\!\! dx\frac{x^{2n}\left(\pi ^2{+}x^2\right)}{x^2{+}\left(\Gamma/T\right)^2}\frac{1}{\cosh^2(x/2 )},\nonumber\;n=0, 1.\\
\mathcal{J}&=\int^{\infty}_{-\infty} dx\;\frac{x^4+4 \pi^2 x^2+3 \pi^4}{x^2+\left(\Gamma/T\right)^2}\;\frac{1}{\cosh^2(x/2)}.
\end{align}
To arrive at the Eq.~\eqref{ym10}, we integrated out the $t$ variable exactly by using the method of contour integration on the following integrals
\begin{align}
\int^{\infty}_{-\infty}& dt\frac{ e^{ixt}}{\cosh^5(\pi T t)}=-\pi\frac{x^4+10 (\pi T)^2 x^2+9 (\pi T)^4}{24(\pi T)^5\;\cosh(x/2T)},\nonumber\\
\int^{\infty}_{-\infty} &dt \;e^{ix t}\frac{\sinh(\pi T t)}{\cosh^4(\pi T t)}=-\frac{i \pi x }{6 (\pi T)^4}\frac{x^2+(\pi T)^2}{\cosh(x/2 T)},\nonumber\\
\int^{\infty}_{-\infty} & dt\; \frac{e^{i tx}}{\cosh^3(\pi T t)}=-\frac{\pi}{2(\pi T)^3}\frac{x^2+(\pi T)^2}{\cosh(x/2T)}.
\end{align}

The general behavior of the transport integrals in Eq.~\eqref{ym10} can be studied by the numerical integration of Eq.~\eqref{nn1}. Nonetheless, the most interesting behavior of the transport integrals are obtainable by an exact analytical procedure. Since the integrals in Eq.~\eqref{nn1} posses two parameters $T$ and $\Gamma$, we define two asymptotic regimes $T\gg\Gamma$ and $T\ll\Gamma$. To calculate analytically the asymptotes of $\mathcal{P}_0$ and $\mathcal{J}$ in the regime of $T\gg\Gamma$, we use the Lorentzian approximation to the delta distribution: $\delta(x)=\lim_{a\to 0} \frac{a}{\pi}\frac{1}{x^2+a^2}$ to obtain
\begin{equation}
\left.\frac{\Gamma}{T}\;\mathcal{P}_0\right|_{T\gg\Gamma}=\pi^3,\;\;\left.\frac{\Gamma}{T}\;\mathcal{J}\right|_{T\gg\Gamma}=3\pi^5.
\end{equation}
The similar result for $\mathcal{P}_1$ is obtained, however, by expanding the corresponding integrand in the Taylor series with respect to $\Gamma/T$ and using the Sommerfeld integrals
\begin{equation}
\left.\mathcal{P}_1\right|_{T\gg\Gamma}=\int^{\infty}_{-\infty}\!\! dx\frac{x^2+\pi^2}{\cosh^2(x/2 )}{+}\mathcal{O}\left(\frac{\Gamma}{T}\right)^2\!\!=\frac{16}{3}\pi^2.
\end{equation}
In addition for the calculation of $\mathcal{P}_1$ in the regime $T\gg\Gamma$, we also expanded the logarithmic factor in Eq.~\eqref{nn1}. Exploiting these procedures, we obtain the following analytical asymptotes of the transport integrals 
\begin{align}\label{danda1}
\left.\mathscr{L}_1\right|_{T\gg\Gamma}&=-\frac{8\pi G_{\rm L}}{9}\frac{T^2}{\ec}\ln\left(\frac{\ec}{T}\right)|r|^2\sin(2\pi N),\nonumber\\
\left.\mathscr{L}_0\right|_{T\gg\Gamma}&=\frac{\pi^3G_{\rm L} T}{8\gamma \ec},\;\left.\mathscr{L}_2\right|_{T\gg\Gamma}=\frac{\pi^5G_{\rm L} T^3}{16\gamma \ec}.
\end{align}
From the transport integrals presented in Eq.~\eqref{danda1}, we obtain the expressions of thermopower and Lorenz ratio 
\begin{align}\label{danda2}
\left.S\right|_{T\gg\Gamma}&=-\frac{64\gamma}{9\pi^2}\ln\left(\frac{E_{\rm C}}{T}\right)|r|^2\sin(2\pi N),\\
&\left.R\right|_{T\gg\Gamma}=\frac{3}{2}-\frac{3}{\pi^2}\left(\left.S\right|_{T\gg\Gamma}\right)^2.
\end{align}
We note that the thermopower Eq.~\eqref{danda2} \DK{(originally uncovered in Ref.~\cite{matv3})} diverges for vanishingly low-temperature which is due to the assumed asymptotic behavior $T\gg\Gamma$. However, thermopower vanishes at the middle of Coulomb blockade valley $N=1/2$ which corresponds to the exact PH symmetric point. At this PH symmetric point, the Lorenz ratio becomes an universal number $R=3/2$ and hence the electronic thermal conductance acquires the quadratic temperature scaling behavior $\left.\mathcal{K}\right|_{T\gg\Gamma, N=1/2}\propto T^2$.

\begin{figure}[t]
\includegraphics[scale=1]{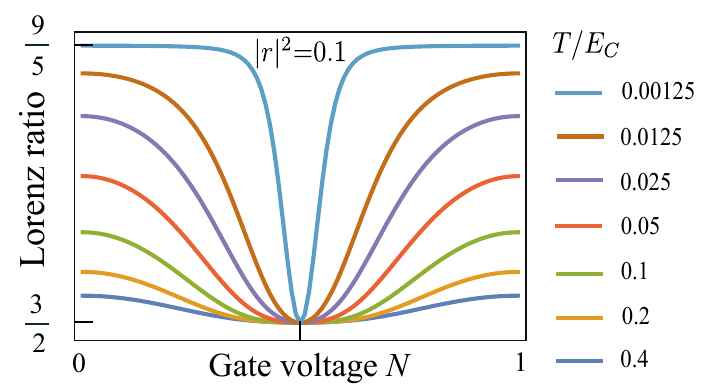}
\caption{Variation of the Lorenz ratio with gate voltage at the strong-coupling regime $T\ll\Gamma$ of the charge-2CK effect for fixed temperatures.}\label{lrfig}
\end{figure}

To obtain the correct behavior of thermopower at very low temperature regime, in the following we consider the opposite limit $T\ll\Gamma$. We use the similar technique as exploited for the discussion $T\gg\Gamma$ regime, namely we Taylor expand the integrand of the integrals $\mathcal{P}_{0, 1}$ and $\mathcal{J}$ given in Eq.~\eqref{ym10} with respect to the small parameter $T/\Gamma$ and retained the lowest order term. Then followed by the use of Sommerfeld integrals, we obtain the asymptotic forms of the transport integrals at low-temperature regime satisfying $T\ll\Gamma$
\begin{align}\label{rama1}
\left.\mathscr{L}_1\right|_{T{\ll}\Gamma}&{=}-\frac{\pi^7 G_{\rm L}}{60\gamma^2}T\left(\!\frac{T}{E_{\rm C}}\!\right)^3\!\!\frac{1}{|r|^2}\frac{\sin(\pi N)}{\cos^3(\pi N)}\!\ln\left(\frac{\ec}{\Gamma}\right)\!,\nonumber\\
\left.\mathscr{L}_0\right|_{T{\ll}\Gamma}&{=}\frac{2G_{\rm L}\pi^2}{3\gamma \ec\Gamma}T^2,\;\;\;\left.\mathscr{L}_2\right|_{T{\ll}\Gamma}{=}\frac{2G_{\rm L}\pi^4}{5\gamma \ec\Gamma} T^4.
\end{align}
Therefore thermopower and the Lorenz ratio in the limit of $T\ll\Gamma$ read~\footnote{This expression of thermopower was predicted in Ref.~\cite{matv3}. The contribution of our work is therefore the analysis of heat transport, namely the behavior of Lorenz ratio in the presence of charge Kondo correlation.}
\begin{align}
\left.S\right|_{T{\ll}\Gamma}&={-}\frac{\pi^3}{5}\frac{T}{\ec}\tan(\pi N)\ln\!\Big[\!\frac{\pi^2/8\gamma}{|r|^2\cos^2(\pi N)}\!\Big],\\
&\left.R\right|_{T{\ll}\Gamma}=\frac{9}{5}-\frac{3}{\pi^2} \left(\left.S\right|_{T{\ll}\Gamma}\right)^2.
\end{align}

\begin{widetext}
\begin{center}
\begin{figure}[h]
\includegraphics[scale=1]{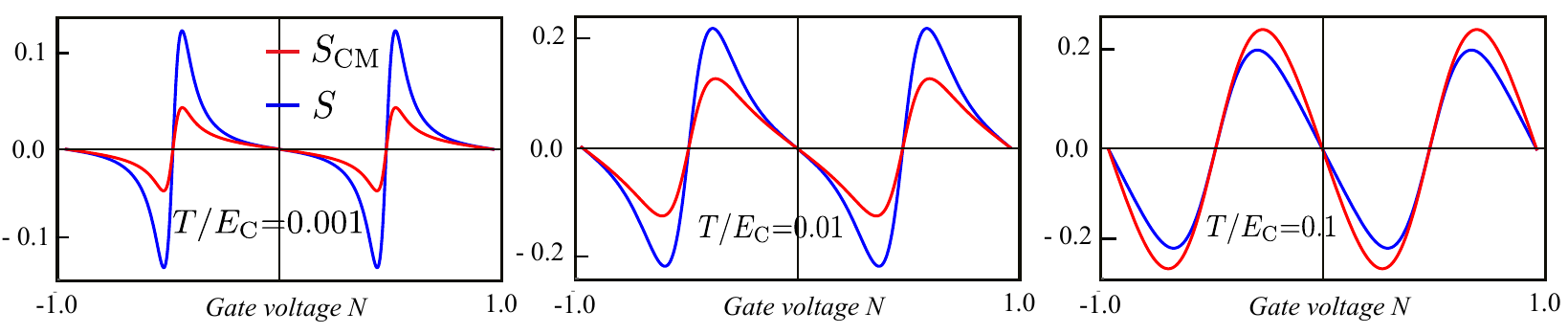}
\caption{Gate voltage variation of thermopower for the charge-2CK effect obtained from the Cutler-Mott relation $S_{\rm CM}$ and by the explicit calculation of thermopower $S$ at different temperatures as indicated in the corresponding plots. For each plots we chose the reflection amplitude to be $|r|^2=0.1$.}\label{mottfig}
\end{figure}
\end{center}
\end{widetext}
\vspace*{-5mm}

\Dk{From the last equation it is seen that the the Lorenz ration also attains an universal number at the PH symmetric pint as similar to the single channel situation (for the corresponding discussion in spin-2CK see Ref.~\cite{dbk})}. The beyond asymptotic behavior of the Lorenz ratio is obtained by numerically solving the Eq.~\eqref{nn1} for $R$, which is presented in Fig.~\ref{lrfig} showing the interplay of $R$ between the regimes $T\ll\Gamma$ and $\Gamma\ll T$. The $T^2$ scaling of the conductance and the linear temperature scaling of the thermopower at low-temperature regime $T\ll\Gamma$ results in the cubic temperature scaling behavior of corresponding electronic thermal conductance $\left.\mathcal{K}\right|_{T{\ll}\Gamma}\propto T^3$.

The temperature scaling behavior and complete interplay of other thermoelectric measures such as the power factor $Q\equiv S^2 G$ and the dimensionless figure of merit $ZT\equiv S^2 G T/\mathcal{K}$ (neglecting the phonon contribution to the thermal conductance) can be easily obtained from the equations~\eqref{danda1} and~\eqref{rama1}. \DK{Since the power factor $Q$ depends linearly on the conductance of left contact $G_{\rm L}\ll G_{\rm R}$, the model of charge-2CK suggested in Ref.~\cite{matv3} however suffers from the smallness of power factor from the application point of view. Given that, however, the figure of merit to the lowest order of approximation does not depends on $G_{\rm L}$ thus providing the appreciable value of $ZT$. }

To study the deviation of thermopower from that predicted by the Cutler-Mott relation given in Eq.~\eqref{mmh}, first we express the conductance Eq.~\eqref{ym10} in terms of the polygamma function. In particular, the integral $\mathcal{P}_0$ in Eq.~\eqref{ym10} is expressed as
\begin{align}\nonumber
\frac{\mathcal{P}_0}{4}&{=}\left(\!\frac{\pi T}{\Gamma}\right)^2\!{+}\Big[1{-}\frac{\Gamma}{2\pi T}\psi^{(1)}\!\left(\frac{1}{2}{+}\frac{\Gamma}{2\pi T}\right)\!\!\Big]\!\!\Big[1{-}\left(\!\frac{\pi T}{\Gamma}\right)^2\!\Big],
\end{align}
where the trigamma function is defied as $\psi^{(1)}(y)=\sum^\infty_{n=0}\left(y+n\right)^{-2}$.
For the SET considered in this work, we have the gate voltage $E\equiv 2\ec N$ which results in the Cutler-Mott thermopower $S_{\rm CM}$ \footnote{To obtain the analytical form of the Cutler-Mott thermopower, we first defined the integrals
$
\mathscr{M}_n=\frac{1}{4}\!\int^{\infty}_{-\infty} dx\; \frac{x^{n+2}}{x^2+a^2}\frac{1}{\cosh^2(x/2)}=\frac{1}{a}{\rm Re}\Big[Z_{n+2}({-}i a)\Big]
$ with the function $Z$ satisfying 
$
Z_n(x)=ib_{n-1}+x Z_{n-1}(x)\;{\rm for}\;n\geq 1
$. The constant factors $b_n$ are obtained as $
b_n\equiv\frac{1}{4}\int_{-\infty }^{\infty } dx\frac{x^n}{\cosh^2(x/2)},\;b_0=1,\;b_2=\frac{\pi^2}{3},\;b_4=\frac{7\pi^4}{15}\cdots
$. Then we used 
$
Z_0(x)=\frac{1}{2\pi}\psi^{(1)}\left(\frac{1}{2}+\frac{ix}{2\pi}\right),
\psi^{(1)}(y)=\sum^\infty_{n=0}\frac{1}{\left(y+n\right)^2}
$
}
\begin{align}\label{jamboo1}
&S_{\rm CM}=-\frac{\pi^2}{6}\frac{T}{\ec}\;\frac{d\ln G}{d N}=\Big[\frac{2 \gamma |r|^2 \sin (2 \pi  N)}{3}\times\nonumber\\
&\frac{4{-}\frac{4\Gamma}{\pi T} \psi ^{(1)}\!\!\left(\frac{1}{2}{+} \frac{\Gamma }{2\pi  T}\right){+}\Big[1{-}\left(\frac{\Gamma}{\pi T}\right)^2\Big] \psi ^{(2)}\!\!\left(\frac{1}{2}{+}\frac{\Gamma }{2\pi  T}\right)}{\frac{2\Gamma}{\pi T}  {+}\Big[1{-}\left(\frac{\Gamma}{\pi T}\right)^2\Big] \psi ^{(1)}\left(\frac{1}{2} +\frac{\Gamma }{2\pi  T}\right)}\Bigg],
\end{align}
with $\psi^{(2)}(y)\equiv \frac{\partial}{\partial y}\psi^{(1)}(y)$ being the tetragamma function. The discrepancy between the thermopower calculated by the formula $S=\mathscr{L}_1/T\mathscr{L}_0$ from the Eq.~\eqref{ym10} and that calculated from the Cutler-Mott formula Eq.~\eqref{jamboo1} is as shown in Fig.~\ref{mottfig}. The significant deviation of thermopower at low temperature from that provided by the Cutler-Mott relation signifies the prominent role of strong electron correlation in the system. Similar behavior of thermopower has been reported in the experimental study of single channel spin Kondo effect~\cite{kiselev}.
\vspace*{-3mm}
\section{Conclusions}
\vspace*{-3mm}
We investigated the heat transport through a SET in charge-1CK (FL) and charge-2CK (NFL) regime by extending the original theory of thermopower developed in Ref.~\cite{matv3}. In addition to the electronic conductance and the thermopower, we obtained the compact analytical expressions of Cutler-Mott thermopower and the WF ratio for both the charge-1CK and charge-2CK effects. \Dk{We characterized the different transport regimes of a SET exhibiting the charge Kondo correlations and explored the universal value of WF ratio at the corresponding particle-hole symmetric point}. The exact asymptotes and the temperature scaling behavior of electronic thermal conductance of a SET in charge-1CK and charge-2CK regime have been explored. Significant deviation of thermopower from the semi-classical Cutler-Mott relation has been investigated and explained in terms of the charge Kondo correlations. An interesting direction for future work would be to investigate the heat transport in the presence of charge Kondo correlation considering more complex setups~\cite{tkt3} and to extending the presented calculations to the beyond linear response regime. The investigations of the impact of channel asymmetry on the heat transport in charge Kondo regime also appears to be a valid avenues for future research.
%
\end{document}